\documentclass[runningheads]{llncs}
\usepackage[T1]{fontenc}
\usepackage{graphicx}
\usepackage{amsmath}
\usepackage{subfigure}
\usepackage{xcolor}

\begin{document}
\title{MFiSP: A Multimodal Fire Spread Prediction Framework}
%
%
\author{Alec Sathiyamoorthy \and 
Wenhao Zhou \orcidID{0009-0000-0413-4282}\and
Xiangmin Zhou \orcidID{0000-0002-1302-818X} \and
Xiaodong Li \orcidID{0000-0003-0346-1526} \and
Iqbal Gondal \orcidID{0000-0001-7963-2446}
}
\authorrunning{Alec et al.}
%
\institute{ School of Computing Technologies, RMIT University, Melbourne, VIC 3000,
Australia \\
\email{alec.sathiyamoorthy@proton.me\\ wenhao.zhou@student.rmit.edu.au \\
\{xiangmin.zhou, xiaodong.li
iqbal.gondal\}@rmit.edu.au}\\
}

\maketitle              
\begin{abstract}
The 2019-2020 Black Summer bushfires in Australia devastated 19 million hectares, destroyed 3,000 homes, and lasted seven months, demonstrating the escalating scale and urgency of wildfire threats requiring better forecasting for effective response.
Traditional fire modeling relies on manual interpretation by Fire Behaviour Analysts (FBAns) and static environmental data, often leading to inaccuracies and operational limitations. Emerging data sources, such as NASA’s FIRMS satellite imagery and Volunteered Geographic Information, offer potential improvements by enabling dynamic fire spread prediction.
This study proposes a Multimodal Fire Spread Prediction Framework (MFiSP) that integrates social media data and remote sensing observations to enhance forecast accuracy. By adapting fuel map manipulation strategies between assimilation cycles, the framework dynamically adjusts fire behavior predictions to align with the observed rate of spread. We evaluate the efficacy of MFiSP using synthetically generated fire event polygons across multiple scenarios, analyzing individual and combined impacts on forecast perimeters. Results suggest that our MFiSP integrating multimodal data can improve fire spread prediction beyond conventional methods reliant on FBAn expertise and static inputs.

\end{abstract}
\section{Introduction}

The 2019-2020 Black Summer bushfires in Australia resulted in 33 fatalities, burned approximately 19 million hectares, destroyed over 3,000 homes, and persisted for more than seven months (nhra-understanding-black-summer). In response, the 2020 Royal Commission on National Natural Disaster Arrangements concluded that disasters are increasing in frequency, intensity, compounding effects, and duration compared to historical precedents (black-summer-royal-commission). The Commission emphasizes that enhanced forecasting could mitigate fatalities through optimized resource allocation for containment and evacuation.

Fire Spread Prediction (FSP) aims to forecast wildfire propagation dynamics, yet traditional methodologies remain constrained by data limitations that impede predictive accuracy. Historically, probabilistic fire risk mapping relies on manually generated outputs subject to persistent operational constraints. Furthermore, conventional fire behavior modeling exhibits critical dependencies on the subjective interpretation of Fire Behavior Analysts (FBAns), introducing substantial uncertainty into projections~\cite{miller,miller-tolhurst,mutthulakshmi}.
Emerging alternative data streams demonstrate significant potential for mitigating these limitations. NASA’s Fire Information for Resource Management System (FIRMS) provides near-global active fire monitoring through multisource satellite imagery assimilation. Concurrently, Volunteered Geographic Information has proven empirically valuable for hazard mapping during remote sensing (RS) data gaps~\cite{florath}. Synthesizing these heterogeneous data sources offers a pathway to reduce empirical estimation dependencies in FSP.
Recent advances facilitate dynamic Rate of Spread (ROS) adjustment between data assimilation cycles by integrating RS feeds and social media data streams~\cite{yoo,zhou-2020}. This integrated approach demonstrates quantifiable potential to outperform conventional techniques dependent upon FBAn expertise and static environmental parameters.

To this end, we propose a Multimodal Fire Spread Prediction Framework (MFiSP) through the investigation of the impact of integrating social media data and remote sensing (RS) observations. To enhance temporal consistency in forecasts, fuel map manipulation strategies adapted from Zhou~\cite{zhou-2020} and Yoo~\cite{yoo} were implemented between assimilation cycles. This approach dynamically adjusts fuel maps to ensure that simulated fire behavior in subsequent forecasts maintains alignment with the observed Rate of Spread. MFiSP efficacy is evaluated using synthetically generated fire event polygons depicting fire locations and behavioral characteristics across multiple scenarios. The individual and combined impacts of these simulated fires on forecast perimeters are systematically analyzed and discussed.
Our contributions are summarized as below:
\begin{itemize}
    \item We propose MFiSP, a Multimodal Fire Spread Prediction Framework through integrating social media data and remote sensing observations. MFiSP fully exploits both the remote sensing and social media data in prediction.     
    \item We derive multimodal inputs for the proposed framework by harvesting geotagged Twitter/Google Maps metadata proximal to arterial roads within fire perimeters.
    \item We conduct extensive experiments over synthetic data generated based on a real fire spread application to prove the high effectiveness of MFiSP.
\end{itemize}
\section{Related Works}
This section synthesizes foundational research across three key domains informing our work, including Fire Simulation Models, Rate of Spread Adjustment, and Social Media Fire Detection.

\textbf{Fire Simulation Models.}
Current fire spread prediction employs two primary empirical approaches: level-set methods (e.g., SPARK) and marker-based techniques (e.g., Phoenix RapidFire, FlamMap). While effective for simulating fire progression using fuel loads, topography, and weather data, these proprietary systems lack data assimilation capabilities. Cellular automata approaches~\cite{mutthulakshmi} incorporated suppression tactics but suffered from oversimplified spread mechanics and the inability to model multiple fires. Hybrid models integrating LSTM networks~\cite{li} or Echo State Networks~\cite{yoo} showed promise but required perfect observational data and failed to account for real-world variability.

\textbf{Rate of Spread Adjustment.}
Advanced data assimilation techniques address observational uncertainties in fire prediction. Zhou-Ding's Ensemble Transform Kalman Filter (ETKF) corrected perimeter errors but ignored spread dynamics~\cite{zhou-ding}, while Zhou's RBFNN-enhanced approach incorporated fuel adjustments using Monte Carlo simulations~\cite{zhou-2020}. However, both methods depend heavily on frequent, high-quality remote sensing data - a limitation for early-stage fires. Yoo's ESN-based method demonstrated accurate predictions but requires two perfect observations for initialization, making it impractical for operational use~\cite{yoo}.

\textbf{Social Media Fire Detection.}
Geospatial wildfire detection via social media leverages three methodological paradigms: gazetteer/NER-based toponym extraction~\cite{zander,florath}, activity deviation analysis~\cite{zander-samuels}, and behavioral topic modeling (LDA/GSDMM)~\cite{blei2003latent,yin2014dirichlet}. Despite their efficacy in fire approximation, these approaches confront persistent limitations: ambiguous nominal location references in NER systems, non-standardized spatio-temporal baselines for deviation metrics, and requisite human validation for topic clusters. Social disaster events have been detected by modeling user retweeting behavior \cite{DBLP:journals/eswa/ChenZSL18} and predicted by influential hashtags in microblogs \cite{DBLP:journals/tkde/ChenZCCSZ22}. However, these approaches have limitations when handling explosive fire disasters requiring dense and accurate data for accurate dynamic fire prediction. Recent hybrid techniques~\cite{DBLP:journals/pvldb/ZhouC22} demonstrate enhanced real-time classification capability, though semantic evolution in user-generated content remains problematic. Contrastively, supervised methods (SVM)~\cite{svm} enable automated behavior analysis but demand extensive training datasets, while unsupervised alternatives (GSDMM)~\cite{yin2014dirichlet} offer operational flexibility at the expense of manual cluster verification—revealing a persistent trade-off between automation rigor and analytical precision in social media-driven fire intelligence.

\section{MFiSP: Multi-source Fire Spread Prediction}
\begin{figure}
    \centering
    \includegraphics[width=1\linewidth]{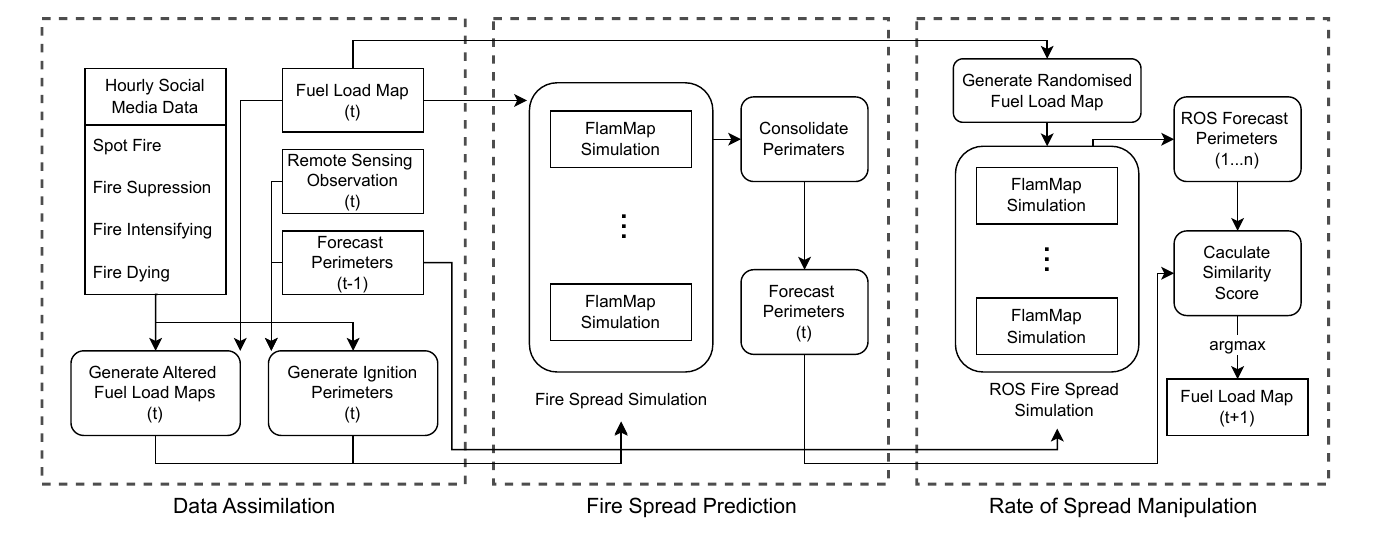}
    \caption{High-level overview of the Multimodal Fire Spread Prediction framework.} 
    \label{fig:framework}
\end{figure}

This section presents the details of the components in our proposed MFiSP framework, illustrated schematically in Figure~\ref{fig:framework}, which integrates three core computational components: (1) Data assimilation generates probabilistic estimates of active fire locations; (2) Fire spread predictions employ Monte Carlo simulations integrating ignition perimeters and fuel map ensembles to project future fire perimeter evolution; and (3) Rate of spread manipulation implements a parameter adaptation framework that formally bridges assimilated fire state characterization with predictive modeling.

\subsection{Preliminary}
Prior to detailing MFiSP's architecture and implementation, we formally define several core concepts fundamental to this paper:
\begin{itemize}
    \item \textbf{Fuel Load:} The quantifiable measure of combustible material within a spatial cell, determining both ignition probability and potential fire intensity.
    \item \textbf{Fuel Load Map:} A geospatial raster layer (typically within a GeoTIFF file) specifying the fuel load value for each individual cell across the landscape.
    \item \textbf{Ignition Perimeters:} Geospatial polygons delineating the initial locations and shapes of active fire fronts from which simulated fire spread originates.
    \item \textbf{Forecast Perimeters:} The resultant set of geospatial polygons generated by executing a fire spread simulation, representing predicted future fire extent.
    \item \textbf{Previous Forecast Perimeters:} The final set of forecast perimeters produced at the conclusion of the preceding data assimilation cycle.
    \item \textbf{Current Forecast Perimeters:} The final set of forecast perimeters produced at the conclusion of the ongoing or most recent data assimilation cycle.
    \item \textbf{Geotag:} Geographic coordinate data (e.g., latitude and longitude) embedded within a non-spatial data source, such as a social media post.
    \item \textbf{Geospatial:} Pertaining to data intrinsically linked to specific geographic locations on the Earth's surface.
\end{itemize}

\subsection{Data Assimilation}
Data assimilation generates probabilistic fire location estimates by integrating inherently unreliable social media data and atmospherically compromised remote sensing observations (Giglio~\cite{giglio}). While valuable for characterizing general fire dynamics (spatial extent, behavior), these inputs lack precision for definitive perimeter mapping. The assimilation pipeline processes hourly SM-derived geospatial polygons representing dynamic fire activity metrics, augmented by periodic RS perimeters when available. Implementation of Zhou~\cite{zhou-2020} and Ding's~\cite{zhou-ding} Ensemble Transform Kalman Filter (ETKF) for real-time adjustment proved prohibitively expensive due to undocumented convergence properties and a lack of libraries for point-level polygon manipulation.

Instead, the methodology adopted a Monte Carlo ensemble approach, as demonstrated by Miller~\cite{miller}. This involved executing multiple fire spread simulations with perturbed input parameters. For each grid cell, the simulated arrival times of the fire front across the ensemble members are recorded. Cells where simulations consistently predicted similar arrival times are assigned higher confidence levels. This ensemble agreement provided a probabilistic estimate of the fire front's likely location at a given time. Although Miller did not utilize assimilation data for forecast correction in the manner attempted by Zhou and Ding, their reliance on the Monte Carlo process to quantify forecast confidence offered a viable strategy for narrowing down probable fire locations. Recognizing inherent inaccuracies within the assimilated data, our approach employ a Monte Carlo process within the forecasting stage to probabilistically predict fire locations, effectively compensating for assimilation uncertainties.

\begin{itemize}
    \item \textbf{Ignition Perimeters.} Geospatially identified previous forecasts are merged with spotting behavior data and RS observations, both subject to inherent inaccuracies. To probabilistically model potential ignition sources and fire spread pathways, numerous fire spread simulations are executed using perturbed environmental parameters (e.g., wind speed/direction, fuel moisture content).
    \item \textbf{Fuel Load Manipulation.} Fire behaviour dynamics—represented as geospatial polygons delineating zones of intensification, suppression, and abatement at specific geographic locations—are used to derive spatially explicit altered fuel load maps. Each altered map is generated by modifying a duplicate of the baseline fuel load map: cells intersecting a given behaviour polygon have their default fuel load values substituted with values corresponding to the represented fire behaviour state (intensification, suppression, abatement). A unique set of altered fuel load maps is produced for every possible combination of the available fire behaviour scenarios.
\end{itemize}
This probabilistic approach accommodates uncertainties in ingested data by leveraging ensemble variability to constrain the most probable fire locations before forecast initiation.

\subsection{Fire Spread Prediction}
Monte Carlo simulations integrate ignition perimeters and fuel map ensembles to predict where the set of active fire perimeters is, based on the information we have assimilated, including the following parts:
\begin{itemize}
    \item \textbf{Ensemble Simulation:} Monte Carlo simulations integrate ignition perimeter data with fuel map ensembles to project fire progression dynamics. The ensemble approach leverages FlamMap's Minimal Travel Time fire growth model, executing simulations under two conditions: an unmodified baseline fuel map scenario and scenarios incorporating advanced ignition perimeters. All simulations model fire spread trajectories over 60-minute intervals using constant wind vector inputs.
    \item \textbf{Consensus Perimeter Identification:} Following simulation completion, all forecasted perimeters are co-registered to a common geospatial reference plane. Spatial consensus is quantified through polygon overlap density measurements per grid cell, establishing hierarchical confidence thresholds: contiguous regions exhibiting above 60\% overlap (designated high-confidence perimeters) are distinguished from areas with lower agreement. The low-confidence perimeter envelope is subsequently derived from the geometric union of all simulated extents where overlap remained below 60\%.
\end{itemize}
The principal advantage of the Monte Carlo approach lies in its comprehensive integration of all assimilation-phase assumptions, which are systematically refined through ensemble filtering to generate current forecast perimeters. These perimeters deterministically define the active fire locations. Upon establishing high-confidence perimeters, the ignition perimeters and altered fuel load maps become operationally redundant and are excluded from subsequent rate of spread calculations.
This high-confidence perimeter serves as the validated fire state for subsequent Rate of Spread Manipulation.

\subsection{Rate of Spread Manipulation}
While previous sections demonstrate successful assimilation of observational data to determine fire locations within each assimilation cycle probabilistically, this methodology does not optimize predictive capability for future fire spread dynamics. To enhance forecast accuracy, we introduce a parameter adaptation framework that calibrates fire growth coefficients against observed propagation patterns.
\begin{itemize}
    \item \textbf{Error Quantification:} 10 random fuel load maps are generated via stochastic perturbation of baseline properties. The fire progression is Simulated from the previous high-confidence perimeter to the currently observed perimeter.  
    \item \textbf{Similarity Scoring:} The spatial overlap ratio between simulated and observed perimeters is computed using Jaccard index:  
    \[ \text{Similarity} = \frac{\text{Area}(S_{\text{sim}} \cap S_{\text{obs}})}{\text{Area}(S_{\text{sim}} \cup S_{\text{obs}})} \]
    \item \textbf{Parameter Selection:} We identify the stochastically perturbed fuel map yielding the highest similarity score. This map is adopted as the baseline for all forecasts until the next assimilation cycle.
\end{itemize}
This transforms raw observational data into actionable model improvements by formalizing the link between assimilation (fire’s current state) and forecasting (predicting its spread). By prioritizing empirical validation (via similarity scoring) and ensemble-based uncertainty propagation, it achieves adaptive forecasting without requiring computationally intensive assimilation in every forecast. 

\section{Experimental Evaluation}

\subsection{Data Collection}
To rigorously assess system performance, comprehensive testing is required. While real-world social media data and hourly fire perimeter observations would have been ideal for validation, the unavailability of such temporally and spatially consistent datasets necessitated the development of synthetic test data. 

\subsubsection{Fire Perimeters}
To address temporal and accuracy constraints inherent in MODIS/FIRMS datasets~\cite{giglio}, synthetic hourly fire perimeters are generated by assimilating geographic fire patterns, behavioral characteristics, and social media reliability metrics. Scenario typology is modeled with parameters derived from 2020 North Complex Fire documentation and 30m-resolution LANDFIRE 2020 inputs. Simulations focus on high-severity WUI zones in California/Colorado (>50\% high-risk classification via FEMA WUI tool), where 2020 wildfires demonstrate extreme behavior. Progression challenges are replicated using ignition points selected through 1-hour test runs with constant 25 km/h winds (225°).

\subsubsection{Social Media Dataset}
Empirical analyses confirm a positive correlation between disaster severity and social media activity intensity~\cite{forati-2021}, with U.S. wildfires yielding substantially more geotagged data than European counterparts, as evidenced by the 2020 Bobcat Fire (236 geotagged tweets) versus the 2022 Landiras Fire (0 geotagged tweets)~\cite{florath}. This geospatial data disparity significantly compromises hazardous area estimation precision. Consequently, two geolocation methodologies are employed: (1) barycenter estimation through tweet density clustering and (2) multi-source information fusion integrating toponyms and infrastructure disruptions derived from textual analysis. Previous barycenter approaches weighted by land cover and population density~\cite{florath,Kitazawa} proved unreliable in mountainous regions, while operational road data~\cite{landfire-2020} inadequately indicate fire origins through closure patterns. To overcome these limitations, synthetic datasets are generated by extracting geotagged metadata from Twitter and Google Maps to establish place-referenced polygons adjacent to arterial roadways within fire perimeters. Four behavioral scenarios are modeled with embedded 12\% geolocation error rates observed empirically~\cite{zander-kumar}. Complementary synthetic remote sensing data comprised complete perimeters sampled from simulations alongside manually degraded observations (30-50\% perimeter omissions) simulating cloud occlusion~\cite{giglio}, sensor obstructions, and temporal data discontinuities.

\subsection{Evaluation Methodology}
We evaluate the effectiveness of our proposed MFiSP by comparing the following alternatives and the existing technique, Partial RS Observation Assimilation.
\begin{itemize}
\item \textbf{Partial Remote Sensing Observation Assimilation. (PR)} Partial RS observations are available from the first hour of the fire and are used to forecast the next hour of the fire. Instead, the previous
forecast is unified with the partial observation into a new set of perimeters. The partial perimeters assimilated are assumed to be of a higher quality than fire behaviour reported through social media. 
\item \textbf{Social Media Data Assimilation (S).} This baseline strategy assimilated hourly social media polygons to generate forecasts evaluated via Otsuka-Ochiai similarity scoring against ground truth perimeters, providing a baseline understanding of the performance of the social media data in fire spread prediction in the absence of RS observation assimilation.
\item \textbf{Partial Remote Sensing Observations and Social Media Data Assimilation (PR + S).} Combining both data streams, this approach first uses social media at hour 1, then merges RS and social media polygons in subsequent hours by geometrically extending overlaps and refining outputs through FlamMap ensembles, achieving synergistic error reduction.
\item \textbf{Remote Sensing Observations and Social Media Data Assimilation (R + S).} Identical to the hybrid strategy but employing full RS perimeters, this upper-bound scenario quantifies how noise-free observations could compensate for social media's inherent geospatial inaccuracies during assimilation.
\item \textbf{Skipped Remote Sensing Observations and Social Media Data As-
similation (Skipped R + S).} With social media ingesting hourly and complete RS data every third interval, this strategy tests forecast resilience to sparse high-quality observations while maintaining continuous crowd-sourced updates.
\item \textbf{Skipped Partial Remote Sensing Observations and Social Media Data Assimilation (Skipped PR + S).} A degraded variant of Strategy 5 using partial RS inputs, evaluating performance when both temporal and spatial data quality are constrained—simulating prolonged cloud cover or sensor outages.
\end{itemize}

\subsection{Experimental Results}
This methodological phase employs three quantitative metrics for experimental evaluation: the Overall Perimeter Similarity Score, the Rate of Spread Similarity Score, and Visualization Results.

\textbf{Overall Similarity Score.}
The Otsuka-Ochiai similarity metric~\cite{romesburg2004cluster} is applied exclusively to assess congruence between terminal prediction perimeters and corresponding empirical fire extents within matching temporal intervals.
\begin{table}[h]
\caption{Overall Similarity Score Results - Final forecast compared to their actual fire perimeters for the same epoch for each assimilation strategy; (S)ocial Media Fire Observations, (P)artial (R)emote Sensing Observations, (R)emote Sensing Observations.}
\label{table:explosive-overall-similarity-score-results}
\begin{center}
\scalebox{0.85}{\begin{tabular}{| p{2cm} | p{1.5cm} | p{1.5cm} | p{1.5cm} | p{1.5cm} | p{1.5cm} | p{1.5cm} |}
\hline
\textbf{Forecast Hour} & \textbf{PR} & \textbf{S} & \textbf{PR + S} & \textbf{R + S} & \textbf{Skipped PR + S} & \textbf{Skipped R + S}  \\
\hline
4 & 0.5783 & 0.6177 & 0.5766 & \textbf{0.7755} & 0.5966 & 0.5609  \\
5 & 0.5093 & 0.6695 & 0.7785 & \textbf{0.8924} & 0.7547 & 0.9001  \\
6 & 0.7026 & 0.8130 & 0.8705 & \textbf{0.9247} & 0.8640 & 0.9000  \\
7 & 0.8730 & 0.8582 & 0.9050 & 0.9160 & 0.9129 & \textbf{0.9291}  \\
\hline
\end{tabular}}
\end{center}
\end{table}
Comparative analysis of assimilation strategies across temporal forecasting intervals reveals significant performance variations in Otsuka-Ochiai similarity scores, as quantified in Table~\ref{table:explosive-overall-similarity-score-results}. 
While standalone Partial Remote Sensing or Social Media fire observations generally produce inferior similarity metrics, their synergistic integration substantially enhances predictive accuracy. This fused approach attained maximum efficacy (92.47 similarity score) at hour 6. Notably, the Skipped S+R strategy remained comparably effective during hours 7-8, consistently surpassing the 92\% similarity threshold.
Crucially, the integrated social media-RS assimilation strategy outperformed all alternatives, including skipped assimilation protocols, which maintained moderate effectiveness. This empirical evidence indicates that synergistic integration of social media data with RS observations significantly enhances predictive accuracy during rapidly expanding or stochastic fire behavior regimes.

\textbf{Rate of Spread Similarity Score.}
Each operational epoch executes 10 fire simulations with perturbed fuel load parameters, progressing from the prior forecast to the current acquisition timing. Resultant perimeters underwent comparative assessment against contemporaneous forecast extents via the Otsuka-Ochiai metric. The optimal fuel load configuration—yielding maximal perimeter congruence—is propagated forward for subsequent simulation cycles, with epoch results representing this maximally concordant iteration. Elevated similarity scores demonstrate improved fuel mapping fidelity relative to forecaste fire behavior.
\begin{table}[h]
\caption{Best of 10 Rate of Spread Forecast Similarity Score - Similarity score achieved by the best Fuel Load Map when their forecast is compared to the current forecast for each assimilation strategy; (S)ocial Media Fire Observations, (P)artial (R)emote Sensing Observations, (R)emote Sensing Observations}
\label{table:explosive-ros-assimilation-score}
\begin{center}
\scalebox{0.85}{\begin{tabular}{| p{2cm} | p{1.5cm} | p{1.5cm} | p{1.5cm} | p{1.5cm} | p{1.5cm} | p{1.5cm} | p{1.5cm} | p{1.5cm} | p{1.5cm} |}
\hline
\textbf{Forecast Hour} & \textbf{PR} & \textbf{S} & \textbf{PR + S} & \textbf{R + S} & \textbf{Skipped PR + S} & \textbf{Skipped R + S}  \\
\hline
4 & \textbf{0.9841} & 0.7584 & 0.7268 & 0.6980 & 0.7399 & 0.7277 \\
5 & 0.7863 & \textbf{0.8608} & 0.8364 & 0.7904 & 0.8162 & 0.7094 \\
6 & 0.6645 & 0.8891 & 0.9093 & 0.9336 & 0.9147 & \textbf{0.9739} \\
7 & 0.8285 & 0.9707 & 0.9580 & \textbf{0.9906} & 0.9580 & 0.9350 \\ 
\hline
\end{tabular}}
\end{center}
\end{table}
Table~\ref{table:explosive-ros-assimilation-score} compares the similarity scores obtained from running the previous forecast with the best of 10 randomly altered fuel load maps between the different assimilation strategies. 
Initial experimental phases (hours 4-5) reveal that standalone remote sensing and social media assimilation maintained high effectiveness, though integrated application underperformed—likely due to information fidelity deficits during fire ignition phases. Contrastingly, multimodal integration achieved peak performance at hours 6-7, with Skipped S+R and full S+R strategies yielding 97.39 and 99.06 similarity scores, respectively. These metrics demonstrate that data fusion progressively enhances predictive accuracy as fires transition from incipient to developed behavioral regimes.

\begin{figure}
    \centering    
    \includegraphics[width=1\linewidth]{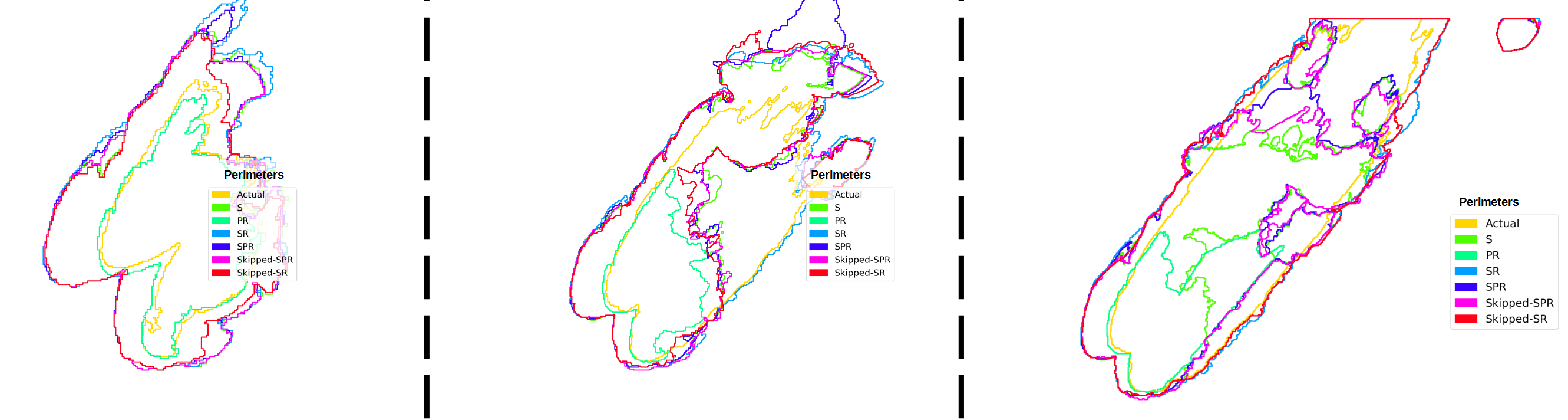}
    \caption{Predicted perimeters compared to their actual fire perimeters during hours 4 to 6 of the explosive case. 4th hour of comparisons (left), 5th hour of comparisons (middle), 6th hour of comparisons (right).}
    \label{fig:predicted perimeters}
\end{figure}
\textbf{Visualization Results.}
Figure~\ref{fig:predicted perimeters} visual comparisons of predicted versus observed fire perimeters (yellow) over 4–6 hours reveal method-specific performance dynamics. At the 4-hour mark, PR delivers the most accurate predictions, aligning with Table~\ref{table:explosive-ros-assimilation-score}. This early-stage advantage likely stems from the smaller, more observable fire footprint combined with inherent latency in social media data propagation. Other methods consistently overpredicted expansion at this stage. By hour 5, concurrent data gaps in both PR and RS observations emerge, enabling fusion-based methods to demonstrate superior predictive capability. Notably, at hour 6, the Skipped-SR method achieves optimal alignment with ground truth, conclusively validating the synergistic integration of social media and remote sensing data streams.

\section{Conclusion}
In this paper, we introduce MFiSP, which synergistically integrates social media data streams and remote sensing observations to significantly enhance wildfire forecasting precision. By dynamically recalibrating fuel map parameters between assimilation cycles, MFiSP achieves unprecedented alignment with observed Rate of Spread metrics. Validation through synthetically generated fire perimeters across diverse behavioral scenarios demonstrates that multimodal data assimilation consistently outperforms conventional forecasting approaches reliant on Fire Behavior Analyst expertise and static environmental inputs. The framework's adaptive architecture establishes a new paradigm for operational fire spread modeling, with particular efficacy in rapidly evolving fire regimes where traditional methods exhibit critical limitations.
In the future, we will focus on applying MFiSP to real-world wildfire datasets to validate the model's performance, generalizability, and operational utility.
%
%

\bibliographystyle{splncs04}

\end{document}